\documentclass[pdflatex,sn-mathphys-num]{sn-jnl}

\usepackage{graphicx}%
\usepackage{multirow}%
\usepackage{amsmath,amssymb,amsfonts}%
\usepackage{amsthm}%
\usepackage{mathrsfs}%
\usepackage[title]{appendix}%
\usepackage{xcolor}%
\usepackage{textcomp}%

\usepackage{manyfoot}%
\usepackage{booktabs}%
\usepackage{algorithm}%
\usepackage{algorithmicx}%
\usepackage{algpseudocode}%
\usepackage{listings}%
\usepackage{subcaption}
\usepackage{hyperref}
\newcommand{\rom}[1]{\MakeUppercase{\romannumeral #1}}

\theoremstyle{thmstyleone}%
\theoremstyle{thmstyletwo}%

\theoremstyle{thmstylethree}%

\begin{document}

\title[SEAMS Phase~\rom{1}]{Space Electric and Magnetic Sensor (SEAMS) Phase~\rom{1}: A Pathfinder Mission for Low-Frequency Radio Astronomy}


\author[1]{\fnm{Atharva} \sur{Kulkarni}}\email{atharvak1994@gmail.com}
\equalcont{These authors contributed equally to this work.}

\author[1]{\fnm{Samruddhi} \sur{Rohokale}}\email{r.samruddhi007@gmail.com}
\equalcont{These authors contributed equally to this work.}
 \author[1]{\fnm{Niketan} \sur{Pawar}}
 \author[1]{\fnm{Raghuttam} \sur{Hombal}}\email{group4pgs@gmail.com}

 \author[1]{\fnm{Nikhil} \sur{Navale}}\email{navale.nikhil@yahoo.in}
 \author*[1]{\fnm{Aditee} \sur{Joshi}}\email{aditee04@gmail.com}

 \author[1]{\fnm{Damayanti} \sur{Gharpure}}\email{dcgharpure@gmail.com}
 \author[1]{\fnm{Subramaniam} \sur{Ananthakrishnan}}\email{subra.anan@gmail.com}
 \author[2]{\fnm{Janardhan} \sur{Padmanabhan}}\email{janardhan.padmanabhan@gmail.com}
 \author[3]{\fnm{Abhirup} \sur{Datta}}\email{abhirup.datta@iiti.ac.in}

\affil*[1]{\orgdiv{Department of Electronic and Instrumentation Science}, \orgname{Savitribai Phule Pune University}, \orgaddress{\street{Ganeshkhind}, \city{Pune}, \postcode{411007}, \state{Maharashtra}, \country{India}}}
\affil[2]{\orgdiv{Physical Research Laboratory}, \orgname{Navrangpura}, \orgaddress{\city{Ahmedabad}, \postcode{380009}, \state{Gujarat}, \country{India}}}
 \affil[3]{\orgdiv{Department of Astronomy, Astrophysics and Space Engineering}, \orgname{IIT Indore}, \orgaddress{\city{Indore}, \postcode{453552}, \state{Madhya Pradesh}, \country{India}}}


\abstract{In this paper, the Space Electric and Magnetic Sensor (SEAMS), a space based mission dedicated to very low frequency (below 16 MHz) observations of the radio sky is proposed. The SEAMS mission is planned in two phases. The first phase is specifically designed for radio frequency interference (RFI) mapping from the low earth orbit (LEO). The low frequency region of the electromagnetic spectrum is filled with a variety of artificial as well as natural RFI. Previous observations have confirmed the effect of this terrestrial interference on astronomical observations. Thus, a map of terrestrial RFI as observed from space would give valuable information to groups working in the low-frequency regime. 
For the first phase, the payload will be mounted on to the fourth stage of the Indian Space Research Organisation (ISRO)'s third generation Polar Satellite Launch Vehicle (PSLV), which will stay in LEO for several weeks. The payload consists of
a pair of orthogonal active monopoles, a simple baseband sampling receiver optimized
for recording power spectra from each antenna. The design also aims to detect strong signals of astronomical origin despite RFI. In the second phase, we intend to send a more sophisticated space payload to orbit the moon and subsequently land on its surface.}




\maketitle

\section{Introduction}\label{sec1}


The low frequency band of the radio window below 20 MHz (also referred to as ultra-low frequency radio window e.g. \cite{2021RAA....21...85C,2023ExA....56..333Y} ) is the least explored band in astronomy. This band encompasses some of the most interesting science cases such as the study of the early universe using highly red-shifted 21$-$cm radiation from the epoch of reionization and dark ages \cite{2004NewAR..48.1029C,2009NewAR..53....1J}. Other compelling science cases include solar and planetary bursts \cite{2007P&SS...55..598Z,2009NewAR..53....1J}, studying the structure of the
Interstellar medium (ISM) \citep{2002ApJ...575..217P,2009NewAR..53....1J}, study of steep spectrum objects such as radio quasars and galaxies \citep{2009NewAR..53....1J}. Despite these interesting science goals, for many decades, limited contributions have been made at very low frequencies. This is because the observations of astronomical sources become progressively challenging at lower radio frequencies. Telescopes at these frequencies often have lower angular resolutions and lower sensitivities because a huge collecting area is required at longer wavelengths. Further, the turbulence in the Earth's ionosphere makes imaging of the radio sky harder. Since the last couple of decades, low-frequency radio astronomy has witnessed a revived interest, with telescopes like the Murchison Widefield Array (MWA) \citep{2005ASPC..345..452M}, LOw-Frequency ARray (LOFAR) \citep{2013A&A...556A...2V}, Square Kilometre Array (SKA)-low pushing the boundaries of technology at lower frequencies. However, there is a fundamental limit on the lowest observable frequency from the Earth's surface because of the ionospheric cutoff frequency. The plasma frequency of the free electron density in the upper layer of the atmosphere cuts off radiation in the range below 5$-$10\,MHz, depending on the time of the day. Although the opacity of the ionosphere is an advantage for terrestrial radio transmissions, it is a hurdle for low frequency radio astronomy. A space-based solution is therefore required to overcome this barrier below 10 MHz.

The first space missions that conducted radio astronomy below 10 MHz included Radio Astronomy Explorers (RAE $-$ 1 and 2) \citep{1975A&A....40..365A}. The data from these missions provided crucial insights into the challenges of doing radio astronomy from space. It made a surprising discovery of the presence of terrestrial radio frequency interference in space, up to the lunar orbit \citep{1975A&A....40..365A}. These missions were also able to produce a map of terrestrial radio emission for the first time at frequencies below $15$ MHz \citep{1973P&SS...21..443H}. It showed that radio transmitters and urban industrial areas can be one of the prime sources of noise that can escape the ionosphere and cause interference for space based observations \citep{1973P&SS...21..443H}. However, the RAE$-$2 observations showed that the far side of the moon benefits from the shielding provided by the moon itself, and hence is the best site for radio astronomical observations \citep{1975A&A....40..365A,2009NewAR..53....1J}
. Since then, several low-frequency lunar missions have been planned or carried out. These include DARIS \cite{DARIS}, OLFAR \cite{OLFAR}, WIND/WAVES \citep{1996GeoRL..23.1287K} and STEREO/WAVES \citep{2008SSRv..136..487B} and the recently launched suite of experiments with Chang' E 4 \cite{2023ExA....56..333Y} such as Netherland-China Low-Frequency Explorer (NCLE) \citep{2024RaSc...5907906K} and Longjiang 1-2 \citep{2023ExA....56..333Y}. Over the past decades, missions like WIND/WAVES, launched in 1994, made important contributions to low frequency observations such as type III solar bursts, co-rotating  interaction regions (CIRs), magnetic clouds etc, in the range 20 kHz to 14 MHz \citep{Wilson2021}. It also showed that the auroral kilometric radiation (AKR) arising from the interaction of charged particles in the solar wind with the Earth's magnetic field was a strong source of radiation in the frequency range of $500$ kHz to $800$ kHz. 

The Space Electric and Magnetic Sensor (SEAMS) is a low-frequency radio telescope designed and developed by researchers from Savitribai Phule Pune University, Indian Institute of Technology (IIT) Indore, and Physical Research Laboratory (PRL) Ahmedabad, with guidance and funding from the Indian Space Research Organisation (ISRO). SEAMS is designed to operate over a wide frequency range of 300 kHz to 16 MHz, first from the Low Earth Orbit (LEO) and later from the lunar orbit. Its wide frequency coverage enables the study of a variety of science cases, including near-Earth sources such as planetary emissions, magnetic reconnection events, and solar bursts, along with sources of cosmological origin. The mission will be carried out in two phases. The first phase, referred to as Phase~\rom{1}, will be dedicated to mapping the terrestrial RFI from LEO, whereas the second phase will explicitly address the science goals.

The Phase~\rom{1} payload will be mounted on the fourth stage of the Polar Satellite Launch Vehicle (PSLV) and will be put in LEO at an altitude of $\sim475$ km from the Earth's surface. The primary goal for this phase is to make a low-frequency map of the terrestrial RFI as seen from LEO. Phase~\rom{1} is also likely to observe the strongest astrophysical phenomena that can be distinguished from the underlying RFI. Recent efforts to access the ultra-low-frequency radio regime ($<30$ MHz) have demonstrated the feasibility of space-based observations and provided initial constraints on the near-Earth radio environment \cite{8741975,2023ExA....56..333Y,2026AJ....171...26H}. Nevertheless, this frequency range remains only partially explored, particularly below $\sim 10$ MHz where ionospheric opacity prevents ground-based measurements. In this context, SEAMS, operating over 300 kHz to 16 MHz, directly addresses the core ultra-long-wavelength regime. 
The objective of phase~\rom{1} is to establish an observational baseline in this poorly characterised band. The outcomes of this initial phase are intended to give information required to design subsequent phase~\rom{2} of the mission development. The data obtained by SEAMS will provide empirical constraints necessary to refine the foreground models and define the system requirements for all future instruments working in the ultra-low frequency regime. By establishing the necessary technical and scientific groundwork, SEAMS enables future missions to transition from exploratory measurements to precise and targeted investigations of the ultra-low-frequency sky.

In this paper, we give a detailed description of the Phase~\rom{1} receiver. Section \ref{payload} gives the configuration of the SEAMS payload. Section \ref{sec:Signals in LEO} deals with the expected signal strengths in LEO and describes the signal analysis and methodology adopted for an active antenna. Section \ref{Front End} gives an overview of the design considerations of active antennas and receiving techniques. Section \ref{backend} describes the details of the digital backend and its features. Section \ref{Antenna Deployment Mechanism} outlines the mechanical design considerations for realising the antenna and deployment mechanism for the mission. Section \ref{sec:TestingAndVerification} describes the testing and verification of the phase~\rom{1} payload and discusses the results. Section \ref{Conclusion} provides concluding remarks. 

\section{SEAMS Phase~\rom{1} Payload}\label{payload}
As in the case of Earth-based observations, RFI is a hurdle for low-frequency radio astronomy even from space. The presence of a strong RFI in a particular frequency band not only renders that band inaccessible for radio astronomy purposes but also can degrade the sensitivity of the receiver in other frequency bands because of higher order harmonics arising from saturation of the receiver. As mentioned in Section \ref{sec1} previous missions like RAE$-$1 (1968), RAE$-$2 (1972), WIND/WAVES (1994) have given insights of RFI observed in space. These missions also pointed out the visibility of other natural terrestrial sources of RFI such as lightning. However, since these measurements, the utility of modern electronics has increased dramatically, which might have lead to a probable increase in the levels of interference. The fourth stage of PSLV (also known as PS4) provides a unique platform to carry out in-orbit scientific experiments for the duration of a few months in LEO. SEAMS payload will intercept terrestrial radio emissions from different parts of the world and the precession of the PS4 orbit around the Earth will enable the Phase~\rom{1} to produce a global map of RFI. 

\begin{figure}[H]
		\centering
		\includegraphics[width=10cm]{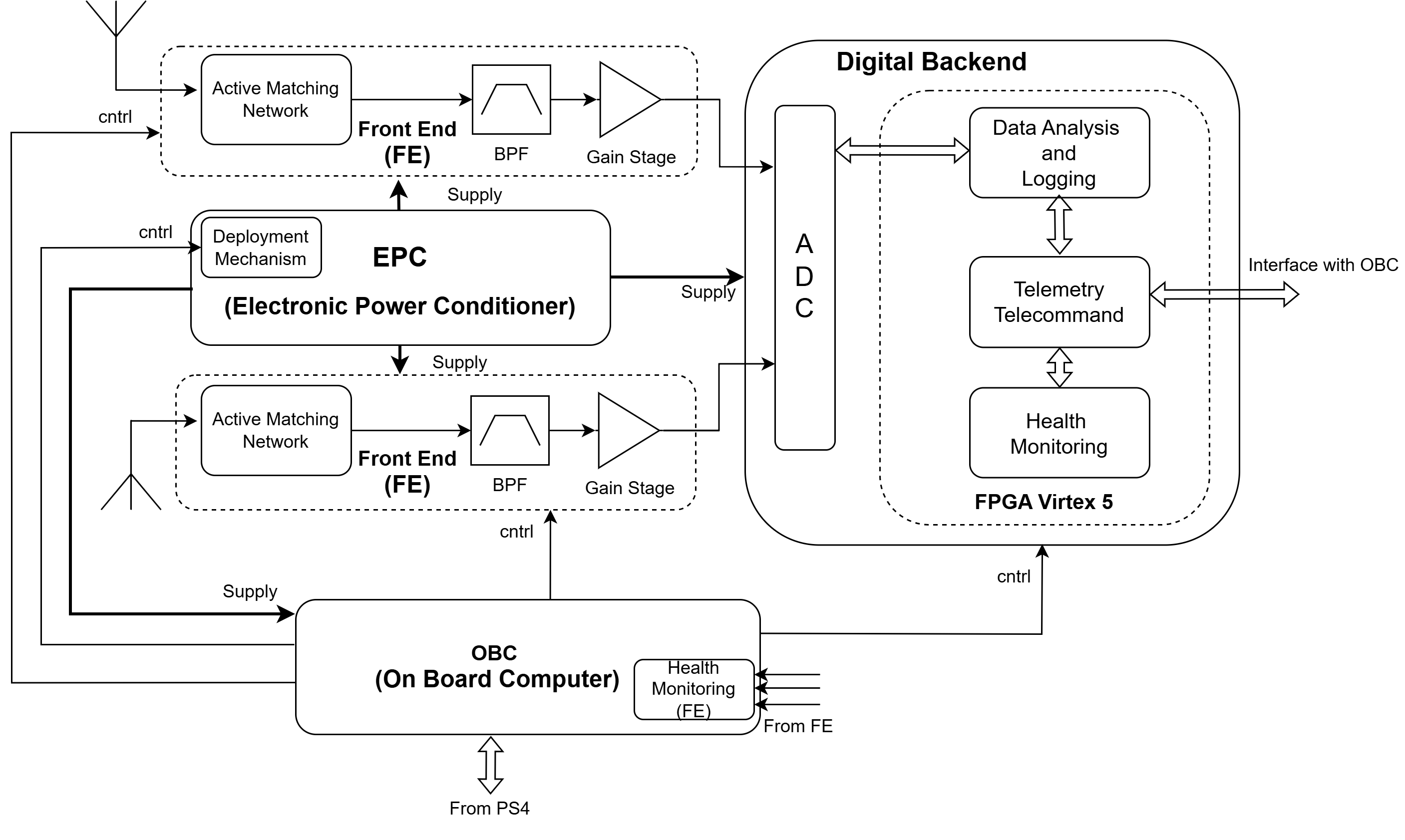 }
		\caption{Functional block diagram of the Phase~\rom{1} payload}
		\label{Proposed Payload}
	\end{figure}

Fig. \ref{Proposed Payload} depicts the functional block diagram of the Phase~\rom{1} payload. A pair of orthogonal active monopole antennas has been used as the electric field sensor. Each active antenna element produces an output that is proportional to the incoming electric field and is impedance matched to the rest of the receiver. The output signals are then filtered and sufficiently amplified by the analog receiver before they are baseband sampled by the digital receiver. The digital data acquisition and analysis system is responsible for producing power spectra corresponding to each channel. The system also includes a solid-state memory interface for logging the acquired data, along with the communication interface for telemetry, telecommand and data transmission. The payload makes use of the PS4's communication link for data transfer to the Earth station. The phase~\rom{1} payload has been developed using commercial off the shelf components (COTS), which makes the system low cost and easy to implement. 
	
It has been shown from previous studies \citep{7771481,Tanti_2023} that the direction of arrival of the radiation can be estimated using a tri-axial electric field sensor. Further, the authors of \cite{7771481} give a case wherein the possibility of detecting the direction of arrival (DOA) of an extended source using a pair of orthogonal detectors is mentioned. While the use of tri-axial electric and magnetic field sensors is planned in the second phase of SEAMS, the success of Phase~\rom{1} might open up a possibility of performing DOA analysis, albeit with very low accuracy, on an extended source in future space missions. Since Phase~\rom{1} is planned with orthogonal monopoles, its design can then be inherited in future applications to incorporate DOA estimation algorithms. To facilitate this, a provision has been made in the first phase to save baseband voltages from both the monopoles. 
Another advantage of using two identical active antennas is that it introduces redundancy in the system which is crucial for space-based applications. Redundancy ensures continuous operation of the payload in the event of a partial failure, where one of the two active antennas may become nonoperational.

\section{Signals in Low Earth Orbit}\label{sec:Signals in LEO}
In this section we discuss the radio environment in LEO. Both astrophysical signals and terrestrial interference occupy the radio frequency spectrum in this range. SEAMS Phase~\rom{1} is intended to map terrestrial RFI from orbit while also leaving a window for detecting powerful astrophysical events. These two scientific goals are described in this section, along with the signal intensity, anticipated sources, and implications for system sensitivity.

The fundamental source of radiation at low frequencies is the non-thermal synchrotron radiation from relativistic electrons primarily in the Milky-way. It forms a background of radiation, known as the galactic background \citep{1978ApJ...221..114N,1979MNRAS.189..465C,2021ApJ...914..128C}. In the absence of other sources of noise, the galactic background limits the sensitivity of the measurements at low frequencies. In this work, we treat the galactic background as a reference to compare the strengths of different signal sources discussed below. We used the empirical model of the galactic background presented by H.V. Canes \citep{1979MNRAS.189..465C}, which accounts for the galactic and the extragalactic components of radio non-thermal radio emissions. 

\subsection{Terrestrial RFI characterisation objectives}\label{Terrestrial RFI}
There are numerous sources of artificial RFI in low frequency radio window such as the communication transmitters, radars, ionosondes, etc. which operate at different frequencies. For simplicity, we divide the sources in two categories: medium-wave (MW) transmitters ($0.5-1.7$ MHz) and short-wave (SW) transmitters ($3-30$ MHz).

\begin{figure}[H]
    \centering
    \includegraphics[width=10cm]{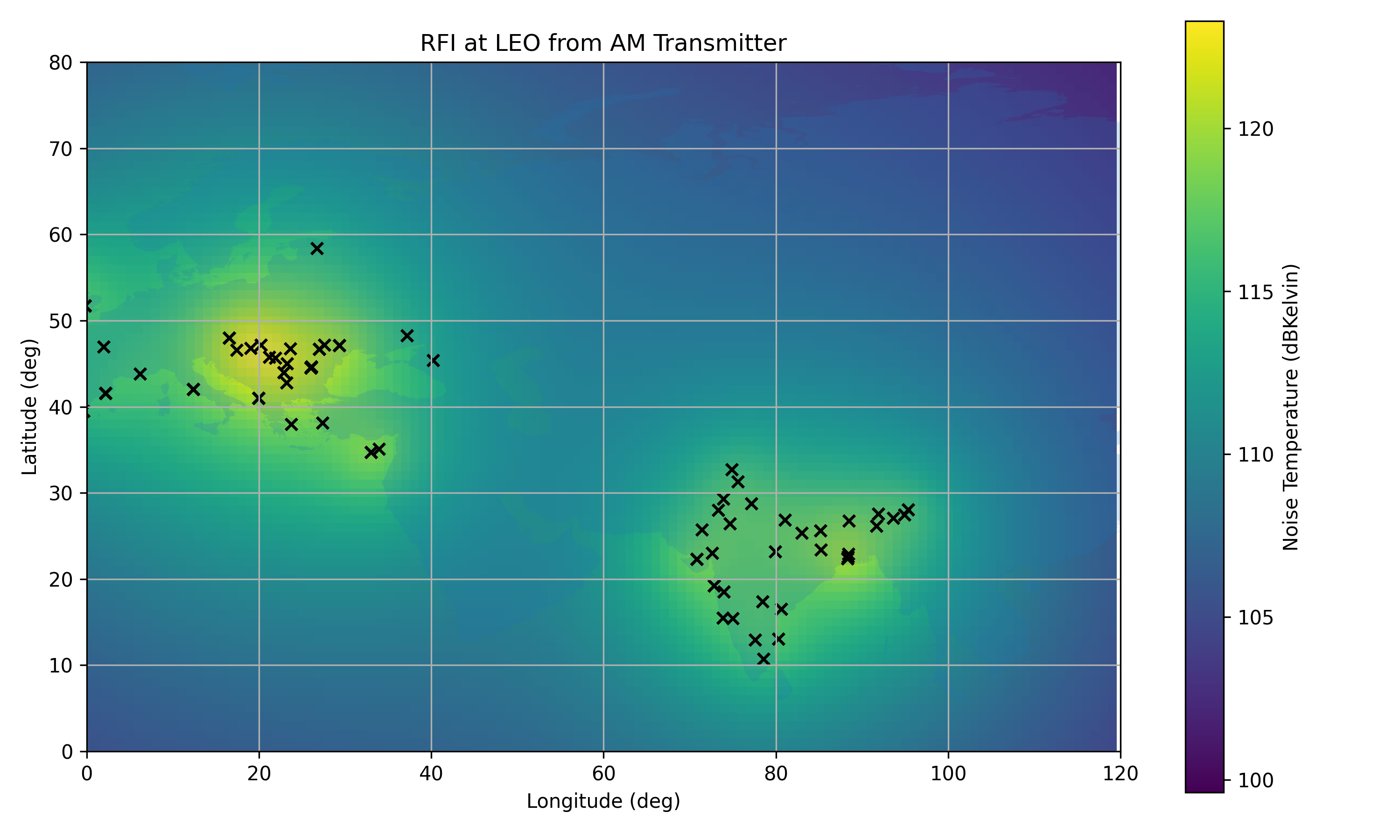}
    \caption{Estimated signal temperature of RFI because of MW radio transmitters across the European continent and India. The black crosses show the MW transmitters that were considered for obtaining the map.}
    \label{fig:RFI_MW_simulation}
\end{figure}

We estimate RFI strength in MW band using the list of medium-wave transmitter stations across India and Europe provided by DRM (Digital Radio Mondiale)  \footnote{\url{https://www.drm.org/wp-content/uploads/2018/09/SCHEDULES-OF-AIR-DRM-MW-A18-oct.pdf}} and MWLIST \footnote{\url{https://www.mwlist.org/mwlist_quick_and_easy.php}} respectively.
We calculate the total power received at the antenna 475 km above the ground by assuming that the transmitter is omnidirectional, the receiving antenna is infinitesimally small and there is only electromagnetic path loss suffered by the radio wave. The total power at any point on the map is the summation of the power arriving at the receiver from every transmitter. The authors of \cite{2017RaSc...52..767Z} describe the ionospheric absorption of HF waves and report a mean total attenuation of 20 dB. They mention that there can be drastic variations in attenuation levels over timescales of a day or even months. Thus to account for the effect of attenuation due to the integrated thickness of the ionosphere or any other atmospheric phenomena, we subtract a mean attenuation of 20 dB from the numbers derived from the simulations. 
Fig. \ref{fig:RFI_MW_simulation} shows a heat map of the estimated power in terms of the corresponding brightness temperature at 475 km above the surface of the Earth\footnote{Since the total power was estimated by adding powers from transmitters at all the frequencies, we calculate the brightness temperature by considering the total receiver bandwidth of 16 MHz}. 

SW transmissions can travel intercontinental distances due to their reflections in the ionosphere. Depending on the plasma frequency of the ionosphere, a significant amount of power from these signals can escape the ionosphere and can be detected from space.  They constitute a major source of interference in low-frequency experiments from space.
To estimate the RFI strength in the SW band, we use previously reported measurements from WIND/WAVES \cite{8741975}, ROLSES$-$1 \cite{2026AJ....171...26H} and Longjiang$-$2 \cite{2023ExA....56..333Y}. Fig. 3 of Bentum et al. (2019) \cite{8741975} shows a comparison of the RFI strength measured at different altitudes during the day and night using the WIND/WAVES instrument. It shows that the RFI is visible with an intensity of $\sim30$ dB above the galactic background at $37,732$km from the Earth's surface.
Similarly, Fig. 9 of Yan et al. (2023) \cite{2023ExA....56..333Y}, shows that Longjiang-2 detected RFI at lunar orbit with an intensity of 2 dB above the galactic background.
Most recently, NASA's lunar mission ROLSES-1 detected terrestrial radio transmissions with power spectral densities (PSDs) of the order of $\sim 10^5 \, \mathrm{nV^2\,Hz^{-1}}$ at a distance of $\sim 3.89 \times 10^5$ km from the Earth \cite{2026AJ....171...26H}. Results from all these missions clearly show that the SW transmitters can escape the ionosphere.

The strength of RFI in LEO at SW frequencies was estimated by scaling the values measured by previous missions to account for free-space propagation losses. The PSD measurements from the ROLSES$-$1 mission were converted to an equivalent brightness temperature within a resolution bandwidth of 16 kHz.
We find that the estimates from the three different missions are consistent with each other and the peak intensity of SW frequencies at LEO is expected to be 50-65 dB above the galactic background. As the measurements from all the missions were recorded well above the ionosphere, the values already include the loss because of ionospheric absorption. Hence we do not adjust for ionospheric attenuation in our estimate at SW frequencies.

\subsection{Opportunities for Astrophysical Detection in LEO}   
Apart from RFI, there can be one-off detections of other natural signals such as Type III solar bursts and planetary emissions from the solar system planets such as the AKR from Earth. The intensity of such events can be in the range of 30-80 dB above galactic background during solar maximum. Then there can be sources that are terrestrial yet natural. For example, lightning in thunderstorms is one of the visible sources in this frequency range, with levels touching 35 dB above the galactic background \citep{1973P&SS...21..443H} depending upon the region on the Earth where it was emitted. 
	
The payload is expected to intercept the F layer of the ionosphere at its designated height. This means that there may exist a narrow window of opportunity for detecting astrophysical signals while transiting over the sunlit side of the Earth, as the RFI from the Earth may be reduced because of increased absorption in the D layer. This creates a possibility of detection of sufficiently strong Type III solar bursts from the sunlit side.
In summary, the Phase~\rom{1} payload is expected to intercept strong RFI from the Earth, and its diurnal variation can be studied. From the day side, there will be absorption from the D layer of the atmosphere, during which there is a possibility of detection of cosmic signals at higher frequencies.

\subsection{Signal Analysis}
	\label{Signal Analysis}
    The strengths of astrophysical signals are often reported in terms of their equivalent brightness temperatures $T_{\rm src}$. Here we derive a framework to relate the source brightness temperature to the equivalent voltage spectral density induced at the terminals of an electrically small antenna. Using the Rayleigh-Jeans approximation, the specific intensity $I_\nu(\nu)$ of a source at frequency $\nu$ is related to the brightness temperature as
\begin{equation}
    I_\nu(\nu) = \frac{2kT_{\rm src}(\nu) }{\lambda^2}
\end{equation}
    For an antenna with an effective area of $A_e(\nu)$, the total available power spectral density from the source is given by 
    \begin{equation}
        P_{\rm src}(\nu) = \frac{1}{2}\int_0^{4\pi} I_{\nu}(\nu) A_{e}(\nu) d\Omega
    \end{equation}
    Here, factor 1/2 takes into account that the power is received only in one polarisation.
    For an electrically small antenna $A_{e}=3\lambda^2/8\pi$ and the shape of the beam is $\Omega$ $\approx$ 4$\pi$. Thus, the integral can be simplified and written as 
    \begin{equation}
        P_{\rm src}(\nu) = \frac{3}{2}kT_{\rm src}(\nu)
        \label{eq;power_to_temp}
    \end{equation}
    If $R_{\rm rad}(\nu)$ is the radiation resistance of the antenna, then the voltage spectral density induced at the antenna terminal corresponding to $P_{\rm src}(\nu)$ is given by 
    \begin{equation}
        V_{\rm src}(\nu) = \sqrt{4P_{\rm src}(\nu)R_{\rm rad}(\nu)}
    \label{eq:voltage_to_power}
    \end{equation}
    Combining equations \ref{eq;power_to_temp} and \ref{eq:voltage_to_power} we get the relation between the source brightness temperature and the voltage spectral density as 
    \begin{equation}
        V_{\rm src}(\nu) = \sqrt{6kT_{\rm src}(\nu)R_{\rm rad}(\nu)}.
    \label{eq:voltage_to_temperature}
    \end{equation}
    For an electrically small antenna,  $R_{rad} \propto l^2$ \cite{Balanis} where $l$ is the length of the antenna, with typical values of the order $10^{-3}\,\Omega$. Equation \ref{eq:voltage_to_temperature}
    highlights the importance of longer antenna lengths (such that the $R_{\rm rad}(\nu)$ is larger) to ensure that the induced voltage spectral density due to the source is sufficiently large. 
    
    In case of an active antenna, a matching network is designed to measure the $V_{\rm src}(\nu)$ induced at the antenna terminals. In this scenario, the strength of the noise generated by the matching network becomes important as it decides the sensitivity of the  receiver. The input referred voltage noise spectral density $V_{\rm rec}(\nu)$ is a useful quantity in this regard. It is a measure of the total noise generated by the matching network that is converted to an equivalent voltage at the input of the network. 
    
    To estimate the sensitivity of the receiver, one method would be to compare the source induced voltage $V_{\rm src}(\nu)$ with the input referred voltage noise of the matching network $V_{\rm rec}(\nu)$. However, in radio astronomy, the receiver characteristics are often reported in terms of system equivalent brightness temperatures $T_{\rm rec}(\nu)$, which can be directly compared with the source brightness temperature $T_{\rm src}$. Thus for simplicity we adopt the second approach by converting $V_{\rm rec}(\nu)$ to an equivalent brightness temperature by using a relation obtained by rearranging the terms in equation \ref{eq:voltage_to_temperature}
    \begin{equation}
        T_{\rm rec}(\nu) = \frac{V_{\rm rec}^2(\nu)}{6kR_{\rm rad}(\nu)}.
        \label{eq:temperature_to_voltage}
    \end{equation}
    The analysis presented in this paper uses equation \ref{eq:temperature_to_voltage} to convert receiver noise to equivalent brightness temperatures, whenever necessary. 

\section{Front end}	\label{Front End}
		\subsection{Active antenna subsystem}

        In addition to observational challenges, there are instrumentation challenges at lower radio frequencies. For example, the size of a resonant antenna at longer wavelengths can be of the order of hundreds of meters to a few kilometers. The deployment of such a large wideband resonant antenna in space is an enormously challenging task. Instead, a different approach is adopted, in which an electrically small antenna is used along with a matching network to match the impedance of the antenna with the rest of the receiver. Such an antenna system is called an active antenna. In this section we discuss the characteristics of the orthogonal monopoles and the matching network. 
        
		\subsubsection{Orthogonal monopoles}
\begin{figure}[H]
			\centering
			\includegraphics[scale=0.06]{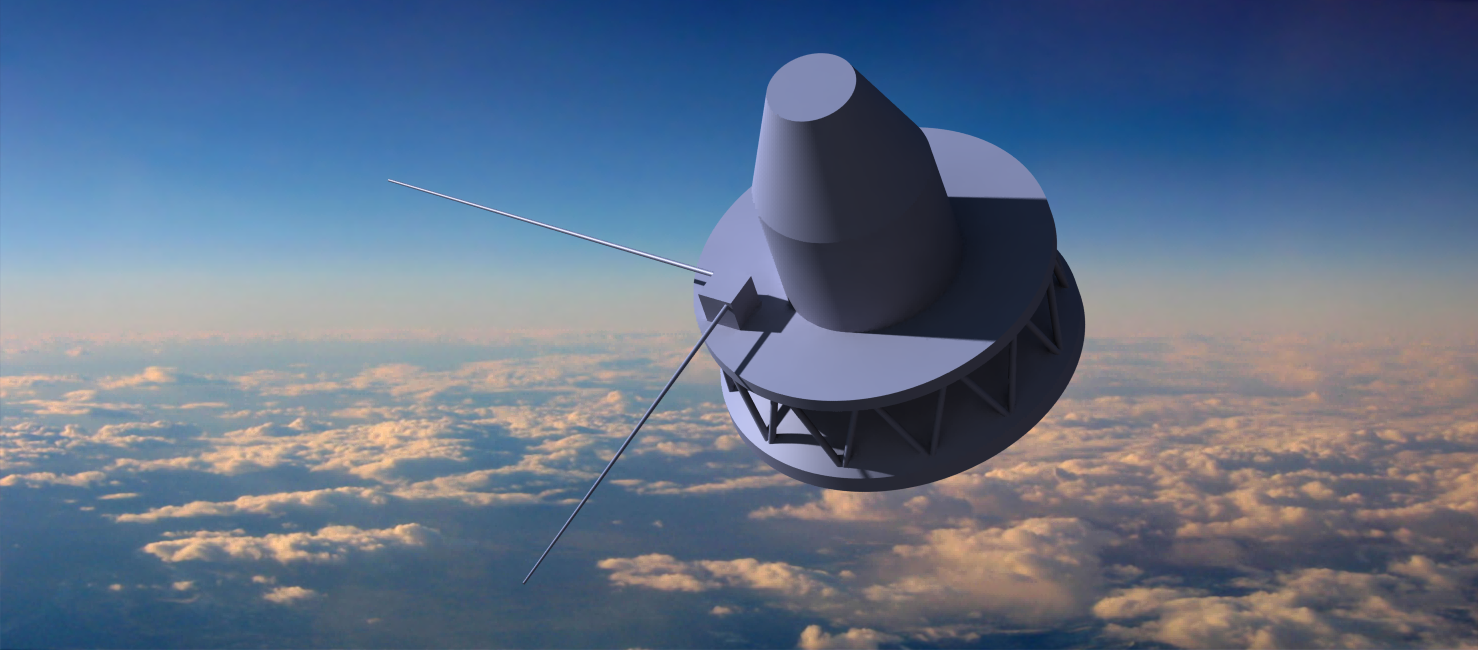}
            
			\caption{The placement and the orientation of the pair of orthogonal monopoles on PS4.}
			\label{fig:Antenna_arrangement}
		\end{figure}

		An active antenna, which consists of an electrically small antenna and a matching network, is a useful option for sensing electric fields at low frequencies.  Among the numerous forms, the monopole antenna is the easiest to build and deploy, particularly in space conditions. Phase~\rom{1} antenna design consists of two 2-meter long monopoles placed orthogonally to detect both polarizations of the incoming signals. 
        The metallic cylindrical body of the PS4 with a diameter of 2.8 m is used as a common ground plane for both monopoles. Once deployed, the antennas will make an angle of 45$^\circ$ to the tangent vector to the cylinder, as shown in Fig. \ref{fig:Antenna_arrangement}. The monopoles are electrically small at the payload’s operating frequency range, and thus are expected to have a doughnut-shaped radiation pattern. Although the monopoles are placed in the close vicinity of the PS4, the antenna properties are not expected to change significantly. This is because the size of the PS4 is very small compared to the wavelengths of EM waves in SEAMS operating range. This was confirmed by performing EM simulations of the monopole antennas mounted on the PS4. The antenna radiation pattern observed from our result closely resembled with that reported in Fig.3b of Yan et. al. (2023) \cite{2023ExA....56..333Y}. The reactance of the monopole antenna is highly capacitive ($C_{\mathrm{ant}}$ $\approx$ 10-30 pF, depending upon the length of the monopole) along with a very small radiation resistance ($R_{\mathrm{ant}}$ $\approx$ 10 of m$\Omega$). Thus, a matching network is required to interface the antenna with the rest of the circuitry. 

		\subsubsection{The matching network}
		The active antenna works on the principle of measuring the voltage or current induced in the antenna element due to incident radiation. This is done with the help of a matching network, which generates an output compatible with the rest of the system. A matching network for Phase~\rom{1} has been designed to measure voltages produced at antenna terminals using a high input impedance amplifier. At low frequencies the sky is dominated by the galactic background, which limits the sensitivity of the telescope. Although it is possible to design galactic background-limited receivers, the main challenge for Phase~\rom{1} is to obtain a high dynamic range so as to measure the strong RFI in LEO while trying to maintain low-noise performance so that signals of astrophysical origin can also be detected within a modest integration time. 
        With these considerations, the RF front end for Phase~\rom{1} has been designed using op-amps with relatively low input referred voltage noise. Op-amps also have the additional advantage of providing a very high input impedance and stability throughout the band.  
		
\begin{figure}[H]
		\centering
		\includegraphics[width=9cm]{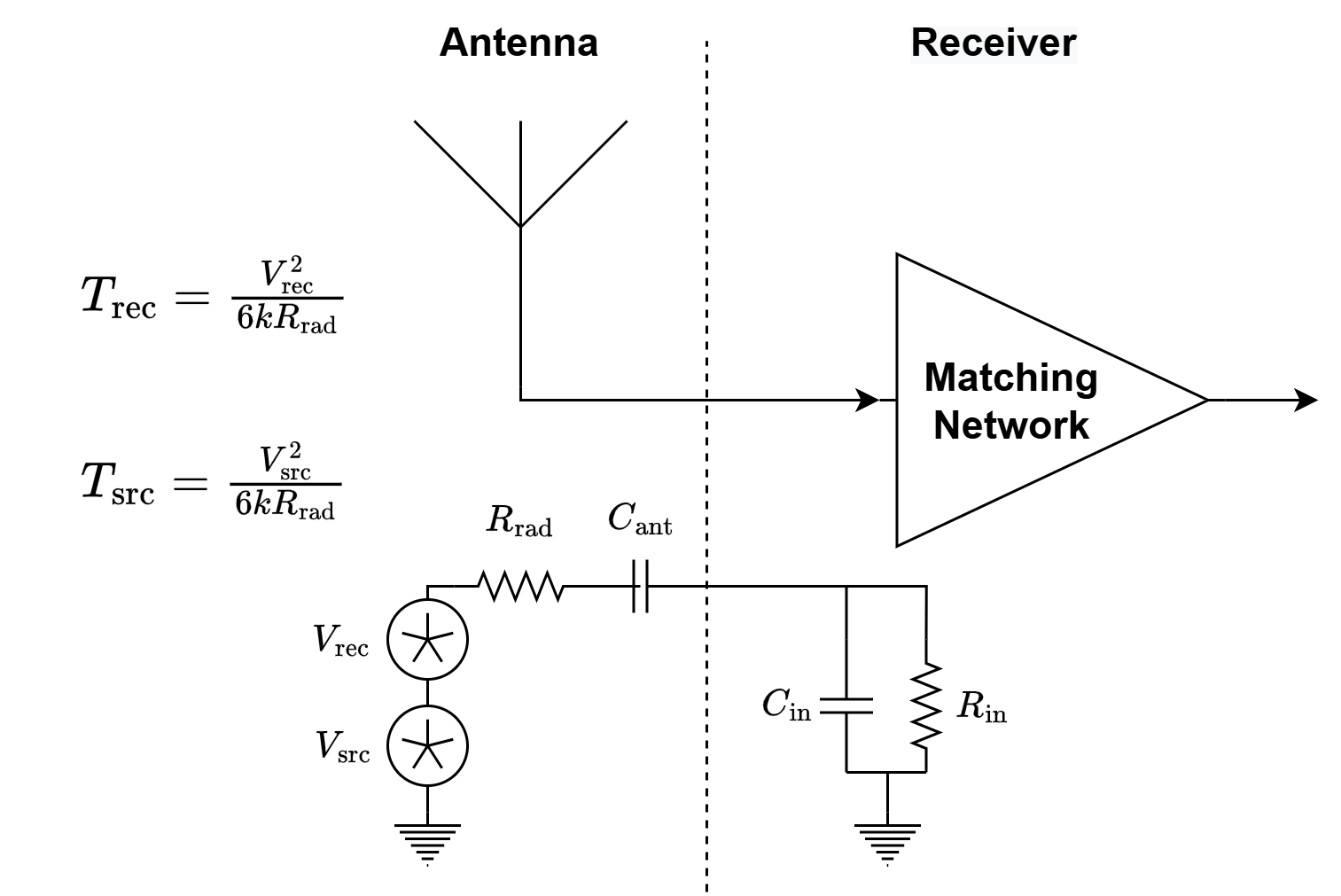}
		\caption{The equivalent circuit of an active antenna}
		\label{Equivalent antenna MN model}
	\end{figure}

		
		The matching network unit for Phase~\rom{1} acts as a buffer and produces a 50 ohm output corresponding to the voltages developed at the antenna terminals. The circuit has a high input impedance of $\approx 10^{12}\Omega$ due to the non-inverting input of the op-amp. Fig. \ref{Equivalent antenna MN model} shows the equivalent circuit of the monopole along with the input stage of the matching network. It can be noted that the voltage  developed at the matching network terminals drop by a factor of $C_{\mathrm{ant}}/(C_{\mathrm{ant}}+C_{\mathrm{in}})$ \cite{2008SSRv..136..529B} due to the capacitive loading as shown in Fig. \ref{Equivalent antenna MN model}. This reduces the minimum detectable flux density of the receiver. The Texas Instruments' transimpedence amplifier OPA653 offers a low $C_{\mathrm{in}}$ of 2.5pF along with low voltage noise of  $6\,\mathrm{nV/\sqrt{Hz}}$, making it the best suited for the application. 
		
\subsection{The filter and the gain stage}
The matching network is followed by a standard 7th order Butterworth LC filter, which restricts the operating range of the payload from 300 kHz to 16 MHz. The Butterworth design ensures that a steep roll-off is obtained in the stop band with minimum ripples in the passband. Another instance of the same design has been used before the ADC as an anti-aliasing filter. This gives an additional advantage of further steepening the stopband roll-off. 
		
A high dynamic range and low distortion gain stage has been used to boost the strength of the received signal. The gain stage consists of an op-amp based low noise inverting amplifier made up of Texas Instruments' LMH6702. This stage also amplifies the noise of the matching network such that the noise floor lies above the minimum detectable level of the ADC. This ensures that the sensitivity of the system is limited by the noise properties of the matching network. Additionally, it is also important to ensure that the gain stage operates in the linear range of the op-amp, as non-linearity would introduce intermodulation products at harmonics of any received radio frequency, making the detection of weak astrophysical signals difficult.
It was experimentally found that a gain of 25 dB is sufficient to adequately amplify the noise while maintaining the linearity of the amplifier. We discuss the details regarding the choice of required gain in section \ref{sec:TestingAndVerification}.

\subsection{Sensitivity and receiver noise temperature}
\label{sensitivity}
\begin{figure}[H]
	\centering
    \includegraphics[scale=0.5]{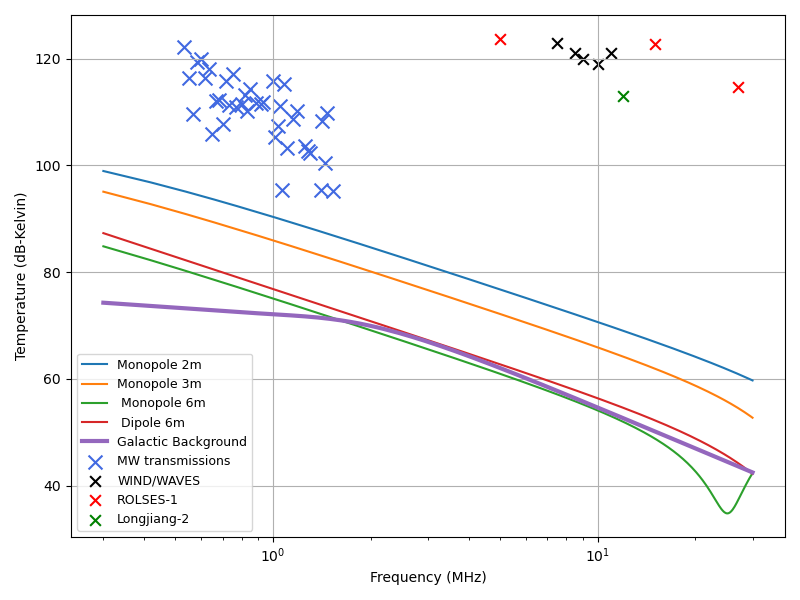}
	\caption{Receiver noise temperature for various lengths of antenna along with the omnipresent galactic background. The blue crosses show the estimated strength of medium wave frequency radio transmitters from Europe and India, as expected in the LEO. An ionospheric attenuation of 20 dB has been assumed in estimating these values (see section \ref{Terrestrial RFI}).  The black, red and green crosses show the estimated RFI based on interpolation of the previous measurements from WIND/WAVES, ROLSES$-$1 and Longjian$-$2 respectively.}
			\label{fig:Sensitivity}
\end{figure}

The two primary elements that determine the sensitivity of the system are the input-referred voltage noise of the receiver and the electrical length of the antenna, as covered in Section \ref{Signal Analysis}.  We estimated the input-referred voltage noise from the measurements of the output-referred noise of the system in a radiation-free setting. The input of the matching network was connected to a monopole antenna, and a Tektronix spectrum analyser was used to track the 50 ohm output. The output-referred noise power of the matching network was found to be $-144\,\text{dBm}/\text{Hz}$
in the passband of the system. This measurement was used to compute the equivalent input-referred noise by taking into account the capacitive loading due to the antenna at the input of the matching network. Although the amount of capacitive loading can be approximated by the ratio $C_{\mathrm{ant}}/(C_{\mathrm{ant}}+C_{\mathrm{in}})$, it may vary due to stray capacitance. Thus for simplicity, we assume a loading factor of $0.8$ and estimate the input-refrred noise to be $V_{\mathrm{rec}}\sim 16\,\text{nV}/\sqrt{\text{Hz}}$.  This value establishes the minimum voltage signal level at which the system can detect any voltage induced at the antenna 
terminals. The receiver temperature ($T_{\text{Rec}}$) concept was utilised to better evaluate 
sensitivity in terms of thermal noise. Equivalent receiver temperature values for 
monopole antennas of various lengths were determined using the computed input-referred 
noise. Both full-wave electromagnetic simulations and analytical methods \cite{Balanis} were used to determine the radiation resistance 
($R_{\text{rad}}$) of these monopoles. The corresponding $T_{\text{Rec}}$ values 
were then calculated using equation \ref{eq:temperature_to_voltage}.
The results from both methods were highly consistent, demonstrating the dependability of the determined receiver temperatures.  Fig. \ref{fig:Sensitivity} shows the calculated $T_{\text{Rec}}$ values for 2, 3 and 6 m monopoles and a 6 m dipole. The anticipated receiver temperatures were also compared to the expected noise levels because of the galactic background, denoted by ${T_\text{sky}}$ in fig \ref{fig:Sensitivity}. 

It is evident from the figure that a longer antenna length should be chosen to obtain a sensitivity limited by the galactic background. However, for the first phase, the expected values of RFI can be in the range 25$-$60 dB above the galactic background depending upon the frequency of the RFI. A longer antenna would increase the sensitivity of the system at the cost of dynamic range. In addition to that, deploying a longer and lighter antenna profile in space on the PS4 platform is an additional challenge. Thus, a $\sim2$ m antenna has been finalized for Phase~\rom{1}, and the deployment strategies have been chosen and set accordingly.
The blue crosses in Fig. \ref{fig:Sensitivity} correspond to the simulated terrestrial MW transmitters from Europe and India that are typically 30-45 dB above the galactic background (see also section \ref{Terrestrial RFI} for discussion on the methodology of simulating the data points). The strength of these narrowband signals is estimated with frequency resolution identical to that of the payload's backend system, which helps in mimicking the response of the payload. A chosen blue cross in the figure might be the result of one or more stations transmitting over frequencies within a bandwidth of 16 kHz.
The SW signals marked in black, red and green are obtained using the measurements from WIND/WAVES reported by Bentum et al. (2019) \cite{8741975}, ROLSES$-$1 reported by Hibbard et al. (2026) \cite{2026AJ....171...26H} and Longjiang$-$2 reported by Yan et al. (2023) \citep{2023ExA....56..333Y}. We expect the SW signal strength estimates to be more uncertain than the MW signal strength estimates, since the attenuation increases as the inverse square of the frequency in the D layer of the ionosphere. Yan et al. (2019) \cite{2021RAA....21...85C} states that below 6 MHz no RFI is seen, ascribing this to ionospheric attenuation. Similar absence of RFI around a few MHz is also seen in Fig. 3 (a) of  Bentum et al. (2019) \cite{8741975} for daytime, while power of RFI increases during the night. Hence, we may observe a much attenuated RFI at low frequencies of a few MHz in the time-integrated spectra, depending on the location of the payload in the orbit.  

\subsection{Calibration of the active antenna subsystem}
\label{calibration}
Calibrating an electrically small antenna below 20 MHz is difficult and remains an active area of research in the low-frequency radio astronomy community. One of the techniques that has been attempted by previous missions includes calibrating using the omnipresent galactic background \citep{2001A&A...365..294D,2011RaSc...46.2008Z,2026AJ....171...26H} for systems that are galactic-background limited. However, as shown in Fig. \ref{fig:Sensitivity}, the phase~\rom{1} design is dominated by receiver noise; therefore, this method cannot be adopted. The primary obstacle for phase~\rom{1} calibration is the lack of a standard external source that is also wideband. However, a few possible approaches are being explored. Here we give a brief overview of some of the methods intended for phase \rom{1} calibration. The details of these methods are deferred to a future publication. 

Ionosondes are one of the terrestrial sources of known power that transmit radio pulses below 20 MHz. These are primarily used to study the structure and variability of the ionosphere directly above them. Thus, most ionosondes transmit signals in high elevation angles and can therefore be detected by the phase~\rom{1} receiver when the payload passes through the ionosonde beams. These detections can potentially be used to calibrate the antenna and the receiver characteristics of the payload. Similarly,  strong solar bursts (Type~\rom{2} and~\rom{3}) or jovian bursts may be detected in the data, which could then offer a way of calibrating the phase~\rom{1} payload. However, the precision and accuracy of the calibration solutions obtained using these methods may be limited by uncertainties in the ionospheric losses at the time of measurement. 
In addition to these, the method of equivalent antenna capacitance substitution can also be used to arrive at a calibration solution \cite{2012nsem.conf..135S,song2014}. This method relies on injecting a known calibration signal into the receiver while replacing the antenna with an equivalent capacitive load at the front-end input. However, one of the limitations of this method is that the true value of the effective antenna capacitance may vary depending on the actual length of the antenna after deployment and the effect of stray capacitance at the interface between the antenna and the matching network \cite{2008SSRv..136..529B}.

\section{Data acquisition and control system}
\label{backend}
\begin{figure}[H]
	    \centering
	        \includegraphics[scale=0.2]{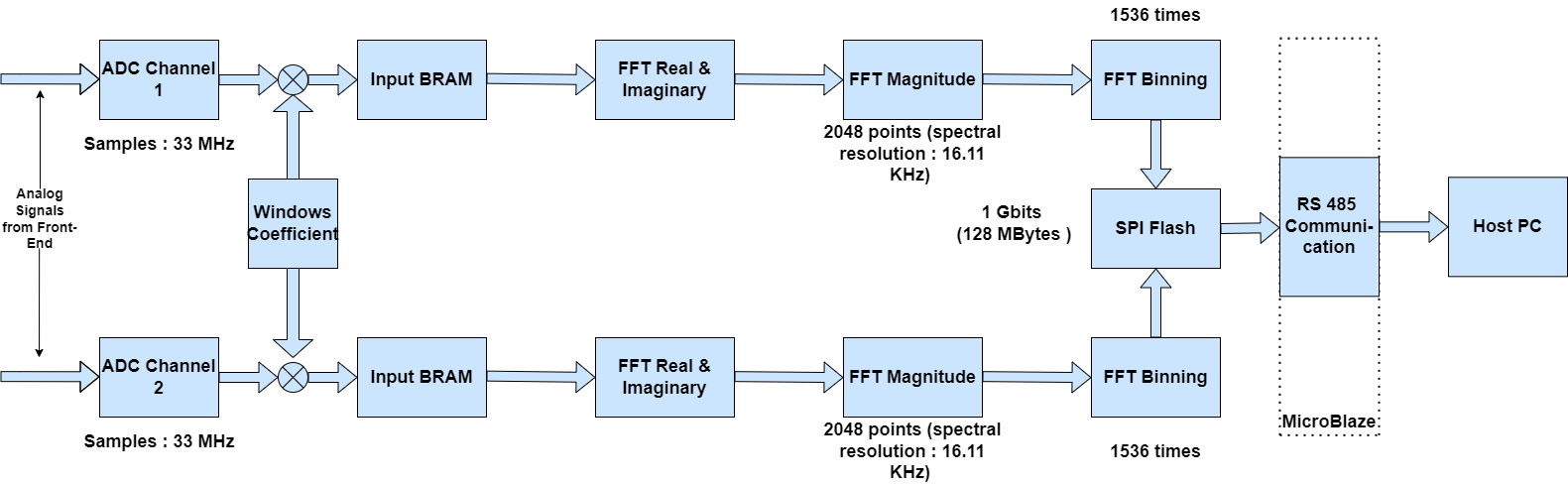}
	        \caption{Digital receiver data flow diagram}
	        \label{FPGA flow diagram}
	    \end{figure}

The digital backend has been envisioned for performing twofold tasks. The first of which is to run the logic control state machine, responsible for analysing the signals and acquiring data on both channels. This data acquisition system is designed in the form of a direct sampling receiver as shown in Fig. \ref{FPGA flow diagram}, implemented using a 12-bit high-speed ADC and Virtex5 FPGA, which samples the signal at the Nyquist sampling rate of 33 MHz. The sampled data is passed through a Hamming window  and stored in the BRAM. A pipeline architecture is implemented to compute a quasi-real-time 2048 point Fast Fourier Transform (FFT) \citep{Borade02012018}. The power spectrum on each channel is computed, with a frequency resolution of $\sim$ 16 kHz and stored in the SPI NAND flash. The RS485 interface has been used to control the system operations and access the data from the NAND flash.
The output produced by the acquisition system is stored in an onboard 1 GB flash memory and is later passed on to PS4 communication link at a suitable time. 

\begin{figure}[H]
	        \centering
	        \includegraphics[width= 8cm]{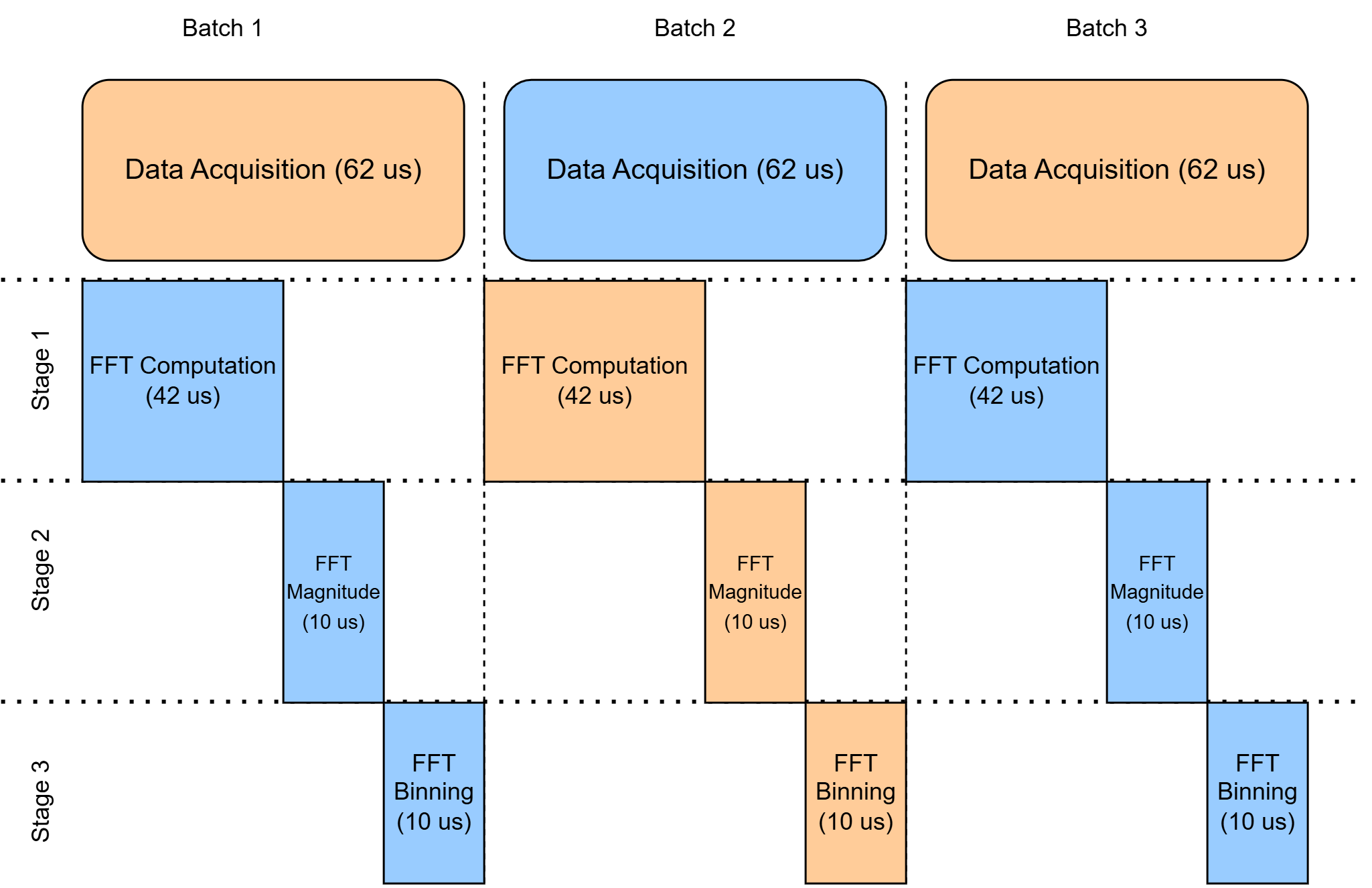}
	        \caption{FPGA pipeline for mode three}
	        \label{FPGA pipeline}
	   
	\end{figure}

The system can operate in three modes. In the first mode, the voltages from each channel are directly saved as 12 bit, 2048 point sampled raw data. This data can be downloaded and analysed offline to obtain a finer spectral resolution. The second mode acquires the complex valued FFT from each channel and stores it in the flash RAM with 16 bit storage for both the real and imaginary components. This data could be used for phase measurements. In the third mode the system acquires time averaged power spectrum of each channel. For this mode, as shown in Fig. \ref{FPGA pipeline}, the optimised pipeline captures and processes raw voltage data every 62 $\mu$s. To reduce the data rate and increase the sensitivity of the measurements, the power spectra from each cycle are integrated over a period of $\sim$ 250 ms. Each acquisition of the integrated spectra from both channels generates 8 kB of data. With this data rate, the system can run continuously for 45 minutes before it exhausts the available space in onboard flash. For testing purpose, the system  control and monitoring is done  with a laptop connected via RS485 interface.

The second task performed by the back-end system involves system health monitoring and providing a command interpreter to control operations of the digital back-end. 
A dedicated On-Board Computer (OBC) based on the Atmega 256 controller is designed for this purpose. The main job of this unit is to synchronise various operations that are required to be performed throughout the duration of the mission. The OBC interprets the commands given by the PS4 telemetry/telecommand interface and responds to it with the appropriate data. To ensure anticipated working, the system performs various housekeeping tasks including monitoring power supply voltages to each subsystem, as well as temperature of the front end and backend. The OBC also generates appropriate signals to control the modules and other actuators during a particular state of operation.

	
	Table (\ref{System Specs}) summarises the characteristics of the Phase~\rom{1} receiver.
		\begin{table}[h!]
		\begin{center}
			\begin{minipage}{190pt}
			\caption{SEAMS Phase~\rom{1} Specifications}
			\label{System Specs}
			\begin{tabular}{p{3.5cm} p{3.5cm}}
				\toprule
				Parameter & Value\\
				\midrule
				Antenna & Pair of Orthogonal Monopoles each of 2 meter length\\
				Beam Pattern & Omni-directional \\
				Matching Network gain & 25 dB\\
				$T_{Sys}$ & 20-30 dB above galactic background\\
				Bandwidth & 16.5 MHz \\
				Resolution Bandwidth & $\sim$16 kHz\\
				ADC  & 12 bits\\
				Integration Time & $\sim$250 ms\\
				Spectral Analysis & Auto Correlation Power Spectra\\
				\botrule
			\end{tabular}
			
		\end{minipage}			    
		\end{center}

		\end{table}

\section{Antenna Deployment Mechanism}
\label{Antenna Deployment Mechanism}
    The design of the antenna profile and its deployment mechanism is one of the crucial aspects of the first phase. Various materials and structures have been simulated and analysed using different materials to realise the antenna and the deployment mechanism. The antenna structure had to be flexible to be stowed in a small volume and have stiffness during deployment, which is a challenging task. Several simulations and experiments were carried out to optimise the stiff and flexible antenna design. Finally, the popularly known STACER profile has been finalised for Phase~\rom{1} \cite{ma13071650}.
    
    An electrically short antenna has been fabricated, with an optimised length of  about 2 m, using stainless steel (SS301 grade). Due to the size and space constraints of the launch vehicle, the antenna element, prior to deployment, has to be stowed in a compact form to be deployed when in LEO, based on command from the ground. 
    
    For the deployment of the STACER, a nichrome wire cutter has been developed after finalising the antenna profile. This deployment system has been patented \cite{Monopole2024}. The storing force required for STACER to remain in undeployed conditions is $\approx$ 23 N. This value is calculated experimentally by adding weights in the opposite direction of deployment. When the STACER is deployed, it generates a jerk, the magnitude of which depends on the stored potential energy in the undeployed state \cite{Stacer_Jiang,Stacer_Pankow}. However, this jerk is found to be negligible compared to the mass of the PS4 platform.  

The natural frequency of the deployment mechanism in the undeployed STACER condition was measured using a vibration shaker machine. It was observed that the natural frequency of the antenna deployment mechanism is 849.09 Hz. It may be noted that when performing a resonance scan, the vibration table exposes the component to controlled forced vibrations through a range of frequencies in an effort to study the component response vibrations. The forced vibration levels are low (typically 1/2 g) and are considered non-damaging. 
 The component response accelerations (CRA), which is given by the output sensor attached to our unit, are compared to the controlled table accelerations (CTA) that comes from the sensor of the vibration table. Any amplification equal or greater than 2 to 1 is typically considered a resonance and the corresponding frequency is called the resonance frequency. In the case of the 2 m antenna, it was observed that the CRA was 18.683 g, while the CTA is 4 g, giving a ratio of 4.670   at a resonance frequency of 849.09 Hz.

\section{Testing and verification}\label{sec:TestingAndVerification}
\begin{figure}
    \centering
    \includegraphics[width=0.5\linewidth]{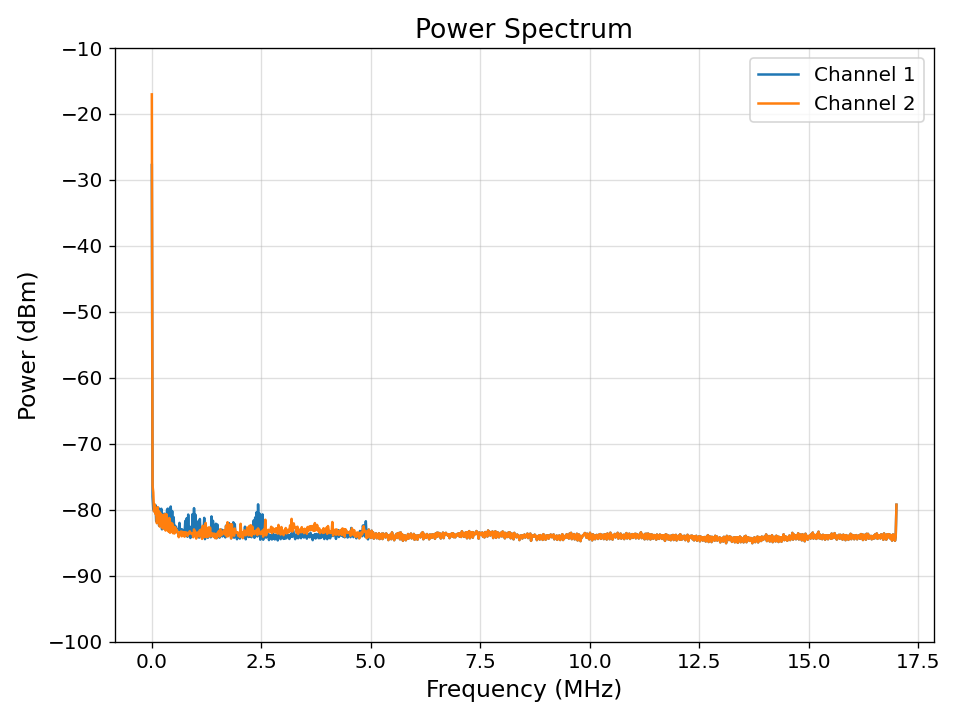}
    \caption{Noise floor of phase~\rom{1} digital receiver.}
    \label{fig:ADC_noise}
\end{figure}

The system was first extensively tested in the lab environment, where the focus was on the verification of its performance and debugging. The initial experiments showed the presence of self-interference at lower frequencies, spanning multiple harmonics. The primary source of this interference was recognized to be the switching signals in the digital receiver that were coupled to the ADC. To mitigate the coupling, the layout of the digital backend board was modified and the switching circuits were shielded. This resulted in a drastic improvement in the noise floor of the system. Fig. \ref{fig:ADC_noise} shows the noise floor of the digital backend when the inputs of both channels were terminated with $50\,\Omega$. It also marks the lowest detectable level of $\sim -84$ dBm for the digital backend when operating with a frequency resolution of $16$ kHz. Using this measurement, we estimated the minimum gain required to amplify the noise of the matching network such that it lies above the minimum detection limit of the ADC. The matching network's output-referred noise of $-144 \, \mathrm{dBm/Hz}$ corresponds to noise level of $-144 + 42.04 \approx -102\, \mathrm{dBm}$ in 16 kHz frequency resolution. Thus a gain of $25$ dB has been designed for the gain stage.  

\begin{figure}[H]
			\centering
			\includegraphics[scale=0.35]{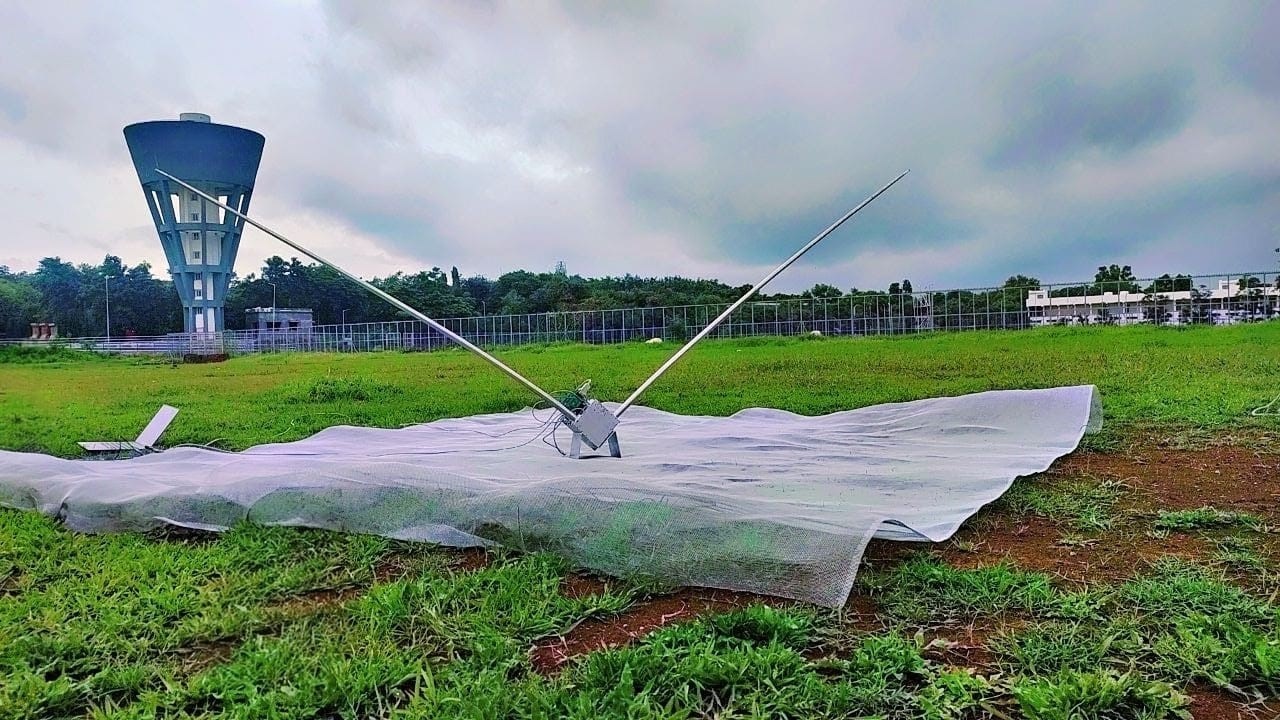}
			\caption{Test setup for SEAMS Phase~\rom{1}}
			\label{fig:Test Setup}
\end{figure}

\begin{figure}[H]
			\centering
    \begin{subfigure}[b]{0.45\textwidth}
        \centering
        \includegraphics[width=\textwidth]{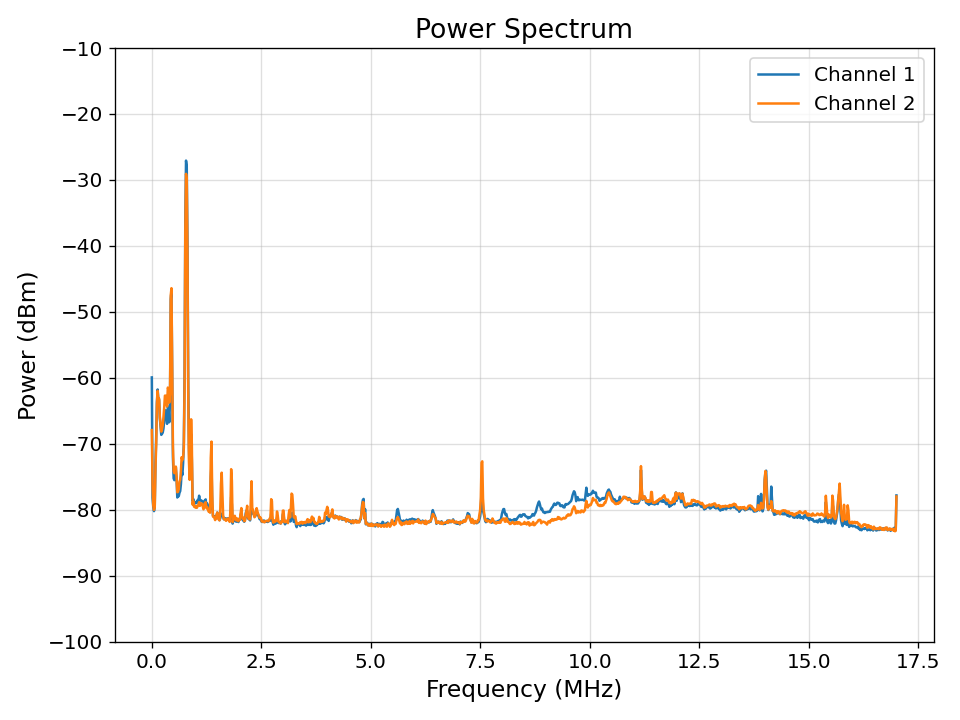}
        \caption{}
        \label{subfig:instru}
    \end{subfigure}
    \begin{subfigure}[b]{0.45\textwidth}
        \centering
        \includegraphics[width=\textwidth]{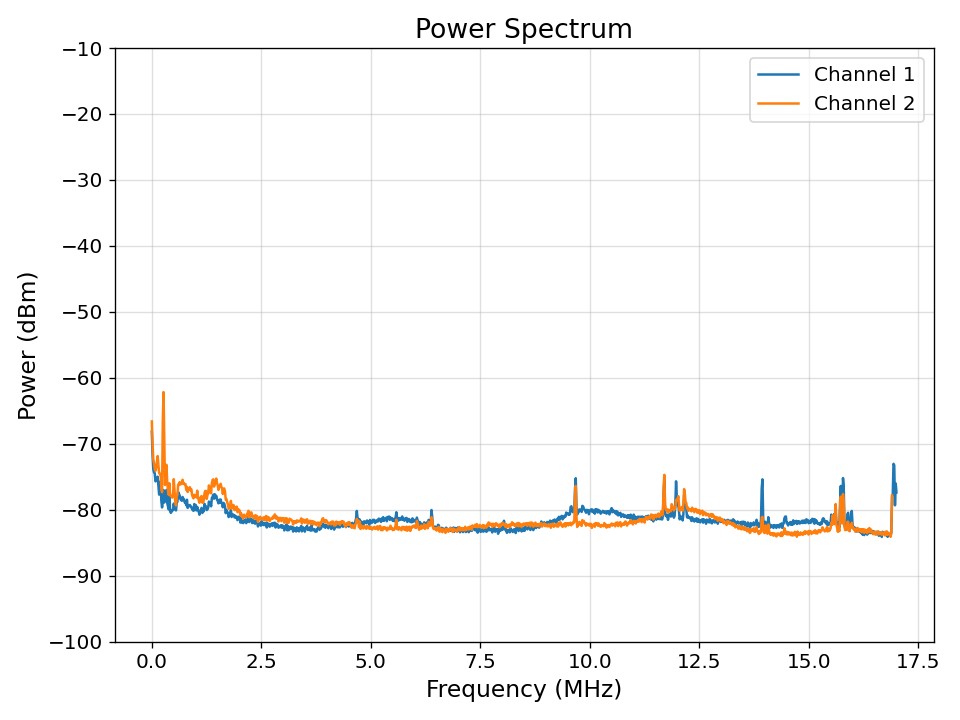}
        \caption{}
        \label{subfig:Dapoli}
    \end{subfigure}
    \caption{ \textbf{(a)} spectra obtained at the Savitribai Phule Pune University ($18.55^\circ$N, $73.82^\circ$E) campus on 18 July 2024. \textbf{(b)} spectra obtained in a rural environment near Dapoli, Maharashtra ($17.76^\circ$N, $73.2^\circ$E) on 23 November 2024.}
    \label{fig:Test_spectra}		
    \end{figure}

Following lab experiments, tests were also carried out on different terrains in varied RFI environments. The cylindrical body of the PS4 serves the purpose of the ground plane for both monopole antennas. To facilitate testing at different locations, we have used a ground plane consisting of a planar aluminium mesh, 3 m × 3 m in size, as a substitute for the complex geometry of the PS4. Fig. \ref{fig:Test Setup} shows the test setup. Each monopole, along with its front end circuitry was enclosed in a metallic box that is placed at the center of the mesh. The effective length of the monopoles for these experiments was 2 meter. The system was powered up using a DC battery. The spectra were recorded in the `average mode'. 
Our initial test setup consisted of a 25 dB gain stage as is designed. However the strength of RFI at certain frequencies in the testing sites was found to be strong enough to saturate the gain stage, causing higher order harmonics in the output. Therefore to avoid saturation, we reduced the gain of the system to 14 dB, only for the purpose of testing the system.
Here we show spectra recorded in three different environments. 
Fig. \ref{subfig:instru} shows the RFI recorded near the Electronics Science Department on the campus of Savitribai Phule Pune University ($18.55^\circ$N, $73.82^\circ$E). This environment is dominated by a local MW radio transmitter operating at 792 KHz that is situated at a distance of $\rm \sim 14$ km from the testing site. This signal can be seen as the strongest spike below 1 MHz in both channels. Additionally, SW radio transmissions near 7.53, 11.16, 14.01, and 15.69 MHz were also detected with very low power. Fig. \ref{subfig:Dapoli} shows spectra obtained in a rural environment near Dapoli ($17.76^\circ$N, $73.2^\circ$E), situated in the western coastal region of Maharashtra. This location is $\rm \sim 110$ km from the Savitribai Phule Pune University and therefore the MW transmission at 792 kHz was not detected. However, SW radio transmissions were still visible at this location.
			          
    \begin{figure}[H]
             \centering
			    \includegraphics[width=\textwidth]{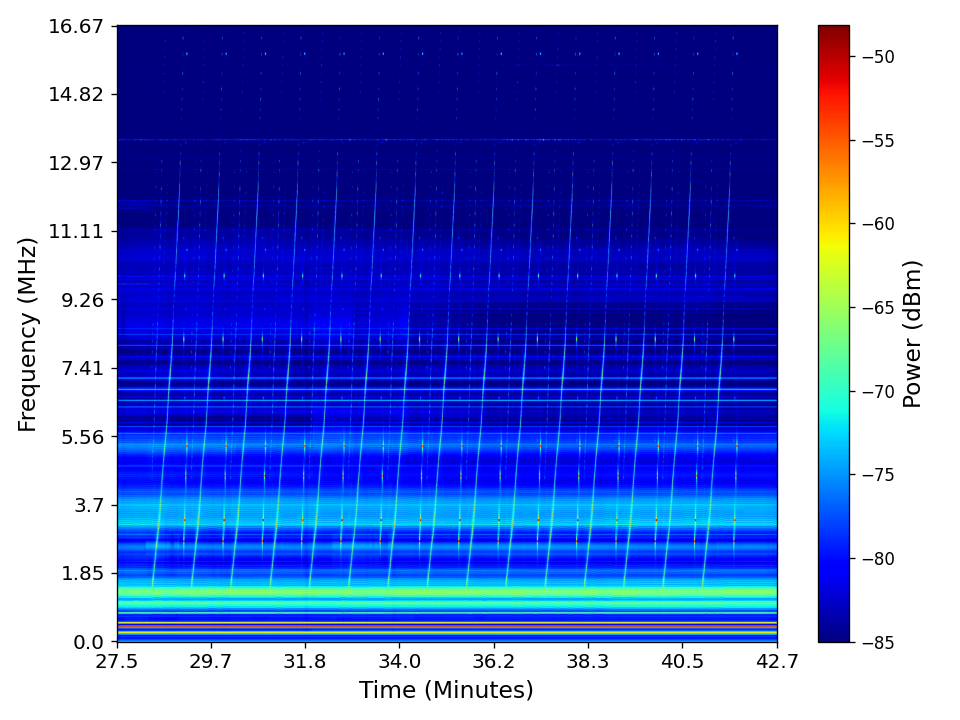}
			    \caption{Spectrogram showing the frequency sweep of an ionosonde at Svalbard, Arctic ($78.92^\circ$N, $11.93^\circ$E) recorded on 10 October 2024 during the Indian Arctic Expedition 2024$-$25. }
			    \label{fig:Chn1_Ionosonde}
		\end{figure}
        
 The plot in  Fig. \ref{fig:Chn1_Ionosonde} shows the spectrogram of the tests conducted in Svalbard, a remote location in the Arctic ($78.92^\circ$N, $11.93^\circ$E). This test was conducted in the vicinity of an operational ionosonde that transmitted a sweep of radio frequencies starting from $1.48$ MHz to $16.32$ MHz, along with a set of four discrete tones at the end of each cycle. Fig. \ref{fig:Chn1_Ionosonde} shows the data collected from one of the channels of the Phase~\rom{1} test payload, where both the frequency sweep and the four discrete tones (red dots) at the end of each cycle can be clearly observed. The details of this experiment will be presented in a separate paper (Joshi et al. in prep).

\section{Conclusion}\label{Conclusion}
The SEAMS project has been largely driven by successive motivated students of the Savitribai Phule Pune University for designing a low cost space payload to be launched in a low earth orbit. Phase~\rom{1} design demonstrates a focused and systematic approach to
characterise the low-frequency radio environment from space, with an emphasis on terrestrial RFI mapping. 
In this paper we describe phase~\rom{1} receiver architecture that has been specially designed for space deployment with a small and reliable front-end system. 
We calculated the expected strength of RFI in LEO using measurements from previous missions such as WIND/WAVES, ROLSES$-$1, and Longjiang$-$2 and arrived at consistent estimates. The front-end electronics were designed using op-amp-based matching networks and gain stages optimized for the expected signal levels. The digital receiver was meticulously designed to provide protection from the self-interference of the payload so that the observations in LEO are not contaminated. The other electronics systems located on the PS4 are also expected to be well shielded; the overall RFI from the PS4 after the payload is integrated remains to be seen 

The system was tested and verified in a variety of RFI environments, including the urban environment at the Savitribai Phule Pune University, a moderately radio-quiet environment near Dapoli and in the vicinity of an ionosonde in the Arctic region. 
The results demonstrate the ability of phase~\rom{1} payload to successfully identify RFI signatures across the payload's operating range. The tests were carried out with a reduced gain of 14 dB; however, as shown in Fig. \ref{fig:Sensitivity}, the strongest expected RFI is about 20-60 dB above the sensitivity of the receiver, and it is well within the dynamic range of the gain stage as well as the ADC used in the digital backend. Therefore, the phase~\rom{1} payload can be operated at its designed gain of 25 dB in LEO. 
At the time of writing this paper, the phase~\rom{1} qualification model is under development and will be used to conduct extreme environment tests such as vibration and thermal tests. 

\bmhead{Acknowledgments}
We thank the anonymous reviewer for their constructive comments and suggestions that helped improve the quality of the manuscript. 
The SEAMS project has evolved over several years and was
sustained by a steady stream of students of the department and
from elsewhere. Firstly, we deeply mourn the loss of Niketan Pawar (a listed
author) in May 2025 in a tragic car accident. His support was invaluable
for the project. He joined as a graduate mechanical engineer but evolved into
an all rounder who conducted the test experiment in the Arctic and other
regions. May his soul rest in peace.
Our special thanks go to Krishna Makhija, Rupali Borade, Kushal Bhalerao, Aditi Nagulpelli, Rasika Sali, George N. George, Prafulla Saste, Kiran Shinde, Viren Mandaogane,
Hemant Patil, Yeshudas Thorat and Harsha A. Tanti and many others before
them. We are deeply grateful to Prof. Avinash Deshpande (ex-Raman Research
Institute, Bangalore) for his insightful advice and critical comments. We thank Prof. Pramod Kale for his support and invaluable advice. We thank
SPPU-ISRO cell as well as the ISRO RESPOND program and ISRO affiliates for their financial and technical support during the past several years. We also thank SCICOM, Pune for their help in designing the backend receiver. JP acknowledges an Indian
National Science Academy (INSA) Senior Scientist Fellowship, and SA, an INSA Senior and Hon. Scientist Fellowship.
Our special thanks go to the scientists from the Tromso Geophysical Group, Drs. Magnar Gullikstad Johnsen and Njål Gulbrandsen, for their help and cooperation for carrying out the experiments near the ionosondes.

\bibliography{references}

\end{document}